# Plasma sheath physics:
# A circuital description, amelioration, and application

Pralay Kumar Karmakar[1], Subham Dutta[1], Utpal Deka[2]
[1]Department of Physics, Tezpur University, Napaam, Tezpur, Assam-784028, India
[2]Department of Physics, Sikkim Manipal Institute of Technology, Sikkim Manipal University, Majitar, Sikkim, 737136, India
Email: pkk@tezu.ernet.in, subham.dtta@gmail.com, utpal.d@smit.smu.edu.in

**Abstract** A synoptic review of the electrical circuital model-based analysis of laboratory plasma sheaths, alongside their stability features in a realistic broader horizon, is systematically presented herein. It explains the basic physics responsible for the inductive ($L_{sh}$), capacitive ($C_{sh}$), and resistive ($R_{sh}$) properties simultaneously, exhibited by plasma sheaths. The analyzed model sheath behaviors are judiciously described in the light of the state-of-the-art sheath scenarios, illustratively. The sheath-based circuital components are minutely contrasted with the traditionally available circuital counterparts. The applications of the novel circuital sheath model in widespread fields of research having both fundamental and applied importance are discussed. The main merits of modelling plasma sheaths through the circuital formalism over the existing non-circuital theoretical ones are briefly outlined, jointly with future applied scope.

## 1 Introduction

The pervasive use of plasmas in diversified industrial and technological applications in daily life is a well-established fact [1]. It is obvious recently that industrial plasma techniques are extensively used in agriculture [2], packaging [3], air cleaning [4], surface hardening [5], plasma thrusters [6], fusion plasma [7], plasma welding [8], plasma medicine [9], etc. It has progressively impacted society in an unprecedented interdisciplinary manner in the essential viewpoint of industrial development and technological growth socio-economically. As a result, the basic insightful physics involved in plasma-wall interaction processes in different physical conditions in the above sectors has been drawing the collective attention of researchers since the broad advent of plasma-based material science, engineering, and technology.

The plasma-based applications mentioned above mainly deal with practically generated laboratory plasmas in different experimental conditions. Besides, plasmas can develop naturally in diversified astrophysical and space environments as well. Although the underlying physics of plasmas is predominantly the same across the environment, the plasmas in different environments vary in terms of basic characteristic differences due to the involvement of various extrinsic source factors, such as the gravitational field, magnetic field, dust constituents, etc. This also results in corresponding atypical applications of studying plasmas in individual environments, as chronologically illustrated and explained in the following.

To begin with, astrophysical plasmas behave as background media for the astro-structure formations. Therefore, studying astrophysical plasmas helps in understanding the formation, life cycle, and death of astro-structures, such as stars, nebulae, galaxies, and so on [10]. The ionospheric space plasma is responsible for radio communication through the reflection of radio frequency waves from the emission antennas to the receiving antennas [11]. Therefore, studying space plasmas helps in communication, navigation, space-weather forecasting, and so forth. The applications of laboratory plasmas are oriented mainly towards applied physics. The applied laboratory plasmas deal with plasma-surface interactions through ionic movement and ion deposition on intended surfaces. Through these interactions, the laboratory plasma system manifests innumerable applications in both material science, electronics, aviation, and so forth [12].

To illustrate the above further, the interaction of plasma and boundary wall intrinsically yields a non-neutral, nonlinear, enveloping intermediate region termed as 'sheath'. The sheath prevents unhindered access of the bulk plasma constituents (electrons, ions, or dust particles) to the chamber wall with a charge separation induced potential difference across its two sides. This bipolar sheath region envelopes the plasma and prevents physical contact with the chamber wall, thereby maintaining charge neutrality across the bulk plasma [13]. This prevents the more mobile thermal bulk electrons from leaking through the chamber's conducting wall. Which would result in a charge number inequality of the bulk plasma electrons and ions, thereby destabilizing it [14].

The sheath region, which hosts plasma-surface interactions, is most relevant in both pure and applied plasma physics in terms of applications. A few of such applications comprise of sputtering through ion energy modulation [15], electron heating through stochastic processes [16], instability excitation through available free energy [17], ion-assisted etching or thin film deposition through suitable potential biasing [18], signal distortion due to perturbation [19], non-invasive plasma diagnosis through spectral analysis [20], plasma sheath wave (PSW) and ion acoustic wave (IAW) formation, various nonlinear effects [21], etc. These applications are elaborately discussed successively through applied circuital model analysis.

The widespread applications of plasma sheath also justify its study through both theoretical and experimental approaches. Therefore, for studying the plasmas across the environments, some compatible theoretical models are used. These models help to understand and anticipate plasmic behavior within different ranges of relevant parametric magnitudes in various plasmic environments.

Such relevant plasma models comprise certain system-specific governing equations dealing uniquely with the individual plasma constituents. These governing equations can be modified as per the presence or absence of various plasma parameters, with corroborating results with the experimental and observational scenarios already reported in the literature [13, 17]. A few of such relevant plasma models are the single particle model, kinetic model, fluid model, multi-fluid model, gyro-kinetic model, etc. [13]. In this article, we discuss plasma sheath in the laboratory setup through another unique plasma sheath model, termed herein as the circuital plasma sheath model [22]. This model application can be justified by the well-known resonant inductor ($L_{sh}$), capacitor ($C_{sh}$), and resistor ($R_{sh}$)-like behavior of the plasma sheath under different applied circumstances as explained ahead.

## 2  Equivalent circuital model

The electrical circuital model is built up to study the Child plasma sheath as an electrical series *LCR*, *CR*, or *LR* circuit with the help of the analytically derived circuital elements ($L_{sh}$, $C_{sh}$, $R_{sh}$) in a systematic manner [23]. The Child current flowing through the sheath circuit can also act as ion implantation current (IIC). This IIC has many applications of great applied value as evident in the literature [22, 23]. Such sheath-equivalent circuital models are known to have extensive applicability in diversified technological areas of research and development.

The lack of analytic closure property of the Child law makes it impossible for the sheath associated IIC and sheath width to be simultaneously evaluated in terms of the applied wall voltage [23]. Conventionally, the Child sheath current is considered equal to the constant Bohm current for evaluating the rest of the parameters. However, no self-consistent physical or mathematical explanation is available so far for the assumed constant current. The circuital model aids to evaluate the self-consistent solution of the Child sheath current and the sheath width in terms of the applied plasma chamber wall voltage and an electrically closed system [22, 23]. The circuital model utilizes the self-consistent Kirchhoff voltage-current law (KVL-KCL) with a single differential equation for mathematical analysis, thereby reasonably corroborating the derived outcomes with existing experimental results [22].

Apart from the above, there are some difficulties with the conventionally utilized non-circuital plasma models, such as the deviation of results from precise outcomes (experimental reporting) at extreme parametric conditions of thermal energy (e.g., the fluid model is only 80% accurate [13]) and high computational cost (e.g., in single particle model, simulations, etc.). In contrast to these conventional plasma models, the phenomenological circuital model works well in non-equilibrium low-temperature plasmas, aimed to study the relevant plasma parameters, such as sustaining voltage, discharge current, dissipated power, space potential, etc. [24], thereby proving its relevance in diversified practical scenarios.

In application, the plasma sheath equivalent electrical circuital model is also applicable for any non-neutral space charge cloud of adequate size with sufficiently excess positive charges. Such clouds comprise of diverse nonlinear structures, such as solitons, double layers (DLs), vortices, and so forth. These charged clouds are produced by excitation and saturation mechanisms during the excitation of plasma instabilities [25].

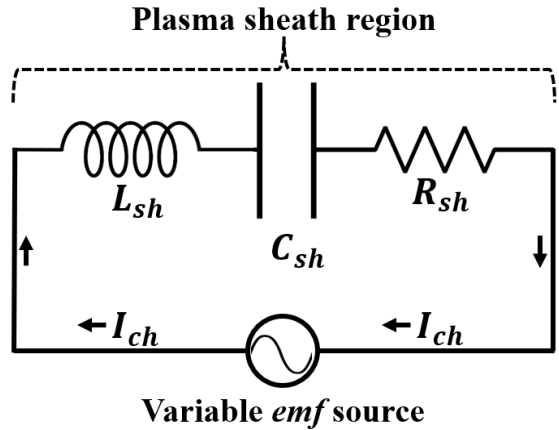

**Fig. 1** Schematic diagram showing plasma sheath behaving as a series *LCR* circuit with a variable *emf* source induced by the temporal variation of the sheath width.

It may be added here that, the analytical expressions for $L_{sh}$, $C_{sh}$, and $R_{sh}$ have been derived in the literature [26], given respectively as

$$L_{sh} = -(c_s \tau_{io}^3 \lambda^2)(40\epsilon_o)^{-1}, \tag{1}$$

$$C_{sh} = \epsilon_o(1 + 0.17\lambda^2)X_O^{-1}, \tag{2}$$

$$R_{sh} = c_s \lambda^2 \tau_{io}^2 (12\epsilon_o)^{-1}. \tag{3}$$

Here, $c_s$, $\tau_{io}$, $\lambda$, $X_o$, and $\epsilon_o$ denote ion sound phase speed, ion transit time through the sheath, characteristic sheath width, equilibrium sheath width, and permittivity of free space, respectively.

We, further, use some experimentally relevant values of various dependent and independent involved parameters, such as the thermal energy, $k_B T_e \sim 2$ eV, equilibrium charge number density, $n_o = 10^{14}$ m$^{-3}$, and ionic mass, $m_i = 1.67 \times 10^{-27}$ kg (for hydrogen plasma) [27]. Using the evaluated term values and expressions, the $L_{sh}$, $C_{sh}$, and $R_{sh}$ values can be further estimated as $-1.36 \times 10^{-7}$ H, $8.84 \times 10^{-8}$ F, and $62.5$ $\Omega$, respectively. These parametric values aid in analyzing the behavior of an oscillating sheath. An oscillating sheath is also known to trigger PSWs in plasmas, elaborated ahead as one of the applications of the circuital plasma sheath model.

Using the above parameters with $X_o = \{1 + 0.17\lambda^2\} c_s \tau_{io}$ and $\tau_{io} \sim 10^{-6}$ s [22], we can express $L_{sh}$ and $R_{sh}$ in terms of the normalized ion plasma oscillation time scale parameter, $\lambda$ as $-3.90\lambda^2 \times 10^{-5}$ and $1.30\lambda^2 \times 10^2$, respectively. After the substitution of $X_o$ in Eq. (2), $C_{sh}$ is found out to be independent of $\lambda$. The parabolic behaviors of $L_{sh}$ and $R_{sh}$ with normalizing factors of $10^2$ and $10^{-5}$, respectively are graphically expressed as

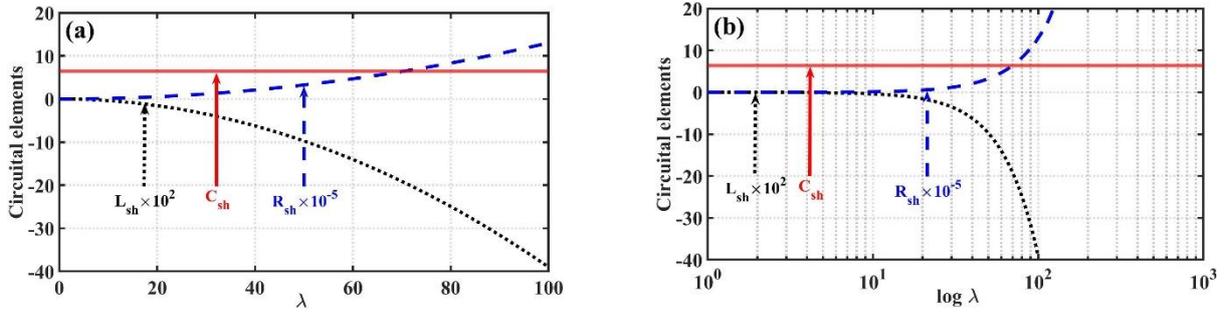

**Fig. 2** Profile of the circuital elements $L_{sh}$, $C_{sh}$, and $R_{sh}$ with $\lambda$ on (a) linear scale and (b) logarithmic scale.

The evaluated negative inductance ($L_{sh} < 0$) value [22] prompts one to explore the possibility and relevance of negative $C_{sh}$ and $R_{sh}$ values also in circuital models. As seen before, these $L_{sh}$, $C_{sh}$, and $R_{sh}$ values can initially be postulated for plasma sheath through circuital model with some characteristic differences to the commercial circuital components. The possible formation of circuital elements and their behaviors in plasma along with both their positive and negative values are discussed below.

It is worth mentioning that a circuit-based plasma model has previously been developed [28], but it differs in several ways from the one proposed in this analysis. In earlier models, the plasma sheath was treated solely as a capacitive component, with its resistive and inductive characteristics neglected. In contrast, the modified circuital model introduced here represents the sheath as having simultaneous inductive, capacitive, and resistive behavior in series, unlike the mixed or non-series configurations of circuit elements used in prior studies [28].

With a brief discussion of the circuital sheath model alongside applied features, we now discuss the basic physical mechanism operating behind the inductive, capacitive, and resistive behaviors of the plasma sheath. We also highlight the typical negative magnitudes of the respective sheath circuital components with self-illustrative schematic and physical description.

## 2.1 Plasma sheath as inductor

Inductors are passive electrical circuit components that store energy in the form of magnetic fields generated by the current flowing through them. The expression for energy stored in the inductor is given by $U_L = LI^2/2$, where $L$ and $I$ are inductance and associated current, respectively. According to the electromagnetic Faraday-Lenz law, a changing current produces a changing magnetic field and a resultant magnetic flux in the inductor. The changing flux induces voltage (*emf*) across the inductor coil in such a way that it opposes its own cause of creation (current and flux change) [29]. Therefore, a positive induction ($L_{sh} > 0$), as mostly exhibited by inductors, indicates their opposition to changes in the current source. The induced electric field through the inductor ($E_L$) and external electric field of the source *emf* ($E_S$), instantaneously act to oppose each other or prevent any change in the source current in accordance with the Faraday-Lenz law of electromagnetic induction [30].

Within a sheath in laboratory plasmas, the undulating, linear, and temporal electric field fluctuates across the sheath produces the fluctuating magnetic field. The field fluctuations occur due to the plasma sheath boundary oscillations resulting in the change in potential gradient, and thus the electric field. This also results in an ion current directional variation. As mentioned above, an ordinary inductor with $L_{sh} > 0$ resists any change in the external electric field, magnetic field, and consequent electric current through it. However, as a special case, a negative inductance ($L_{sh} < 0$) of the sheath may also trigger in. Herein, the induced *emf* across the inductor develops in a way to aid the primary cause of source current fluctuation across the sheath. It implies that any increase or decrease in the current across the sheath continues to self-sustain until a current reversal occurs therein [22].

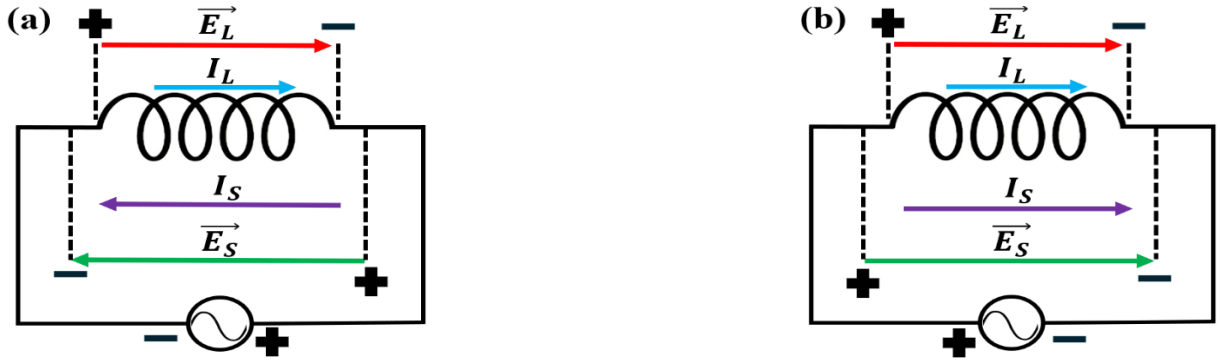

**Fig. 3** Schematic formation of (a) $L_{sh} > 0$ and (b) $L_{sh} < 0$ in plasma sheath circuital system. $\vec{E_L}$ and $I_L$ denote the induced electric field and current in the plasma sheath inductor, respectively. Here, $\vec{E_S}$ and $I_S$ denote the corresponding field and current due to the source *emf*.

The conventional positive inductance occurs when the fields $\vec{E_L}$ and $\vec{E_S}$ act anti-parallel. It occurs due to the Lenz law. The Lenz law can be overcome if the fields $\vec{E_L}$ and $\vec{E_S}$ are parallelly aligned. Such field arrangements positively feedback each other until current reversal.

The $L_{sh} < 0$ can further be illustrated in terms of the changing electric field polarities. If the frequency of $\vec{E_S}$-polarity change is managed in such a way that $\vec{E_S}$ and $\vec{E_L}$ are parallelly aligned, the negative inductance can trigger in. During the $L_{sh} < 0$ condition, both $\vec{E_L}$ and $\vec{E_S}$ align instantaneously parallel towards each other. Due to the temporal variations of field polarities, the frequencies of $\vec{E_L}$- and $\vec{E_S}$-polarity changes must also be equivalent.

In a plasma sheath inductor this atypical behavior can originate due to the availability of free energy sources in the plasma medium in the form of magnetic field, external electric field, thermal energy, etc. The free energy may enable itself to act in aligning $\vec{E_L}$ parallelly with $\vec{E_S}$, thereby, positively feedbacking it, provided its sufficient availability. The electromagnetic energy required to overcome the Lenz law is supplied from the free energy sources.

## 2.2 Plasma sheath as capacitor

It is well known that electrical capacitors are specially designed passive circuit components consisting of a system of conductors (rectangular plates, concentric spherical shells, or coaxial cylindrical elements). These plates store energy in the form of electric fields across them when charged by external power sources. The electrostatic potential energy stored across the plates is given by $U_C = CV^2/2$, where $C$ and $V$ are the capacitance and voltage applied across the capacitor, respectively. The charge in the plates, and hence, the potential energy can be reduced when the external power source is switched off [30].

In a plasma sheath system, the capacitive behavior ideally originates from the bipolar electric charge distribution and the subsequent potential gradient developed across the two sides of the sheath (one along the boundary wall (negative) and other at the plasma sheath junction wall (positive)) comparable to the Debye-scale length. The two sides of the plasma sheath accumulate electric charges of opposite polarities, like that of a parallel plate capacitor (electron-abundant (negative) side and electron-deficient (positive) side). It yields stiff electrostatic potential gradient within a very small spatial distance of the Debye-scale ($\lambda_D$) order. This sheath property recreates a parallel plate capacitor-like arrangement with a plate separation $\sim \lambda_D$ [22]. Therefore, the sheath-plasma capacitor system stores electrostatic energy (in terms of electric field). The intermediate region in between plasma sheath boundaries behaves as an intrinsic dielectric material, thereby enhancing the capacitance, as arranged in commercial capacitors.

A conventional $C_{sh} > 0$ denotes enhanced charge storage with respect to an increase in the electrostatic potential applied across the capacitor plates. The induced electric field across the capacitor plates ($\vec{E_C}$) and the source field ($\vec{E_S}$) are parallelly aligned. Therefore, an increased $\vec{E_S}$ results in a stronger $\vec{E_C}$ and vice-versa.

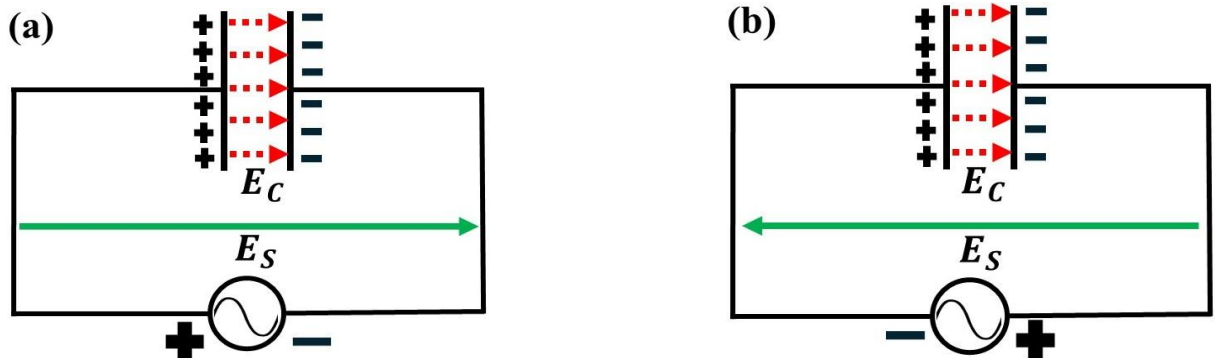

**Fig. 4** Schematic structure of (a) $C_{sh} > 0$ and (b) $C_{sh} < 0$ in plasma sheath circuital system. $\vec{E_C}$ denotes the induced electric field in the sheath capacitor and $\vec{E_S}$, the field of the source *emf*.

For a plasma sheath system, a negative capacitance ($C_{sh} < 0$) denotes reduction of charge content in the plates with the increase in the potential or electric field across them. The capacitor opposes any further increase in charge on its plates. For an alternating *emf* source, during the excitation of $C_{sh} < 0$, the frequency of $\vec{E_S}$-polarity change is equivalent to that of the $\vec{E_C}$ across the capacitor due to the directional variation of current through it. If the $\vec{E_S}$-polarity is varied at a frequency equivalent to the inverse of charging (or discharging) time of the capacitor plates, the $C_{sh} < 0$ can be attained with $\vec{E_S}$ and $\vec{E_C}$ acting anti-parallel to each other. During the $C_{sh} < 0$, an application of external voltage across the capacitor discharges its plates instead of charging them.

In a sheath capacitor, this atypical behavior can originate with some possible availability of free energy in the plasma medium in the form of external electric field, magnetic field, thermal energy, etc. The free energy may enable itself to act in aligning the induced electric field ($\vec{E_C}$) against (anti-parallelly) with the source field ($\vec{E_S}$), thereby nullifying it. It results in a reduction in charge content in the capacitor plates with an increase in *emf*. The energy required to overcome the induced $\vec{E_C}$ is supplied from the free energy source.

## 2.3 Plasma sheath as resistor

A resistor is an electrical circuital anti-flow component. It controls the current by dropping the electric potential across it. In a plasma sheath system, in addition to inductive and capacitive behaviors, resistive behavior is also observed. The electrostatic potential developed across the sheath prevents free flow of charges. This potential barrier (sheath resistor) requires some finite amount of energy to be overcome, and it reduces the net flow of charges through it. This behavior replicates that of a commercial resistor. It is pertinent to add that this plasma sheath resistor differs from a commercial Ohmic resistor in terms of the potential versus current relationship. The potential drop (thus induced *emf*) across the plasma sheath resistor is equivalent to the acquired negative potential on the plasma chamber boundary. The resultant potential gradient also prevents the ions from passing through the stable sheath with velocity less than the Bohm velocity. A spatial variation of the sheath width yields a non-zero temporal variation of the induced *emf* across it [22].

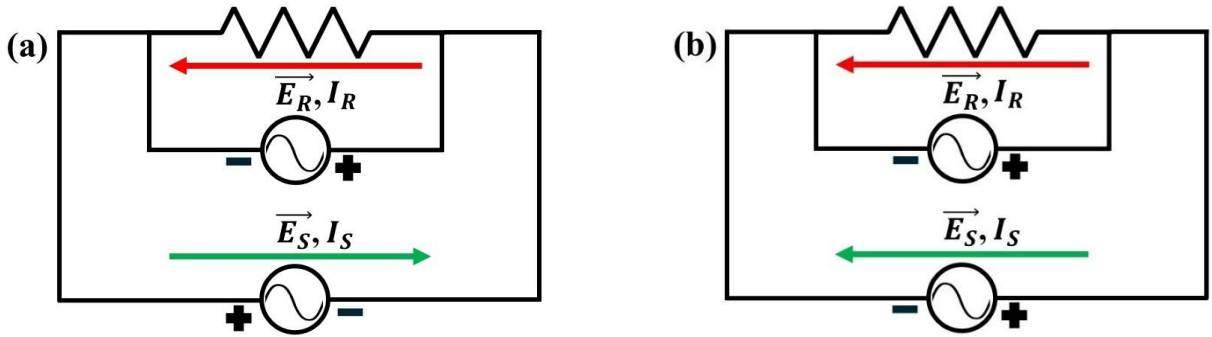

**Fig. 5** Schematics showing possible formation of (a) $R_{sh} > 0$ and (b) $R_{sh} < 0$ in sheath plasma system. $\vec{E_R}$ and $I_R$ denote the induced electric field and current in the plasma sheath resistor, respectively. Here, $\vec{E_S}$ and $I_S$ denote the corresponding field and current due to the source *emf*.

A negative resistance in a non-ohmic plasmic system can arise in the presence of a pre-existing free energy. The initial voltage pulse from an external *emf* source ($V_s$) across the sheath circuit may act as a perturbation (trigger) to reduce the instantaneous current flow in the circuit, thereby yielding an effective negative resistance ($R_{sh} = -dV_s/dI$) therein. This condition can arise in the presence of sufficient free energy than the source *emf*. The free energy overpowers the source *emf* with the latter reducing just to a trigger. Ideally, the plasma sheath circuit behaves as comprising of two individual *emf* sources. One *emf* source yields from the sheath width fluctuation (denoted by characteristics $\vec{E_R}$ and $I_R$) and the other *emf* is induced across the plasma sheath resistor due to the available free energy (denoted by characteristic $\vec{E_S}$ and $I_S$). Since the two *emf* sources are considered to be parallel herein, therefore two apparently anti-parallel electric fields ($\vec{E_R} || -\vec{E_S}$) develop positive resistance ($R_{sh} > 0$), one acting in favor of the other. In contrast, parallel electric fields ($\vec{E_R} || \vec{E_S}$) act against each other yielding a negative resistance ($R_{sh} < 0$). The electric field and the consequent current flow in the smaller circuit (hosting $\vec{E_R}$) is totally nullified and reversed upon activating the larger circuit (hosting both $\vec{E_R}$ and $\vec{E_S}$) with a higher electrostatic potential.

## 3 Phase shift in sheath inductor and capacitor

The magnetic field developed within an inductor prevents any direct passage of current during the first half of the AC pulse due to the Lenz law. The electric current received at the inductor is instantaneously converted into magnetic field resisting its passage (or change) through it. This induced magnetic field yields an instant *emf* (potential drop) across the inductor against the current flow. It results in a temporal delay of current through the inductor. This delay leads to the development of a phase difference of 90° in between the voltage (leading parameter) across the inductor and the current (lagging parameter) through it in any electrical circuit.

It occurs exactly opposite in a capacitor, where both component plates hold some pre-existing free electrons and a gap (either empty or filled with dielectrics) in between them. The physical gap between the plates makes it impossible for voltage and current to change simultaneously. The first pulse of voltage applied across the capacitor gets instantaneously nullified by the relocated free electrons from the plates producing the first pulse of current (leading parameter) before the voltage (lagging parameter) surge. This yields a phase difference of 90° in between the current and voltage parameters. The electrons flow to the *emf* source through the connecting wire but cannot move through the gap amid the capacitor plates.

In addition to the above, an Ohmic resistor is so developed that there is no gap across this resistor as in the case of a parallel plate sheath capacitor, or no magnetic field (an acting Lenz law) as in the sheath inductor to prevent any direct increase in the current. This property aids in forming spontaneous electric fields across it and current through it upon external application of *emf*. Therefore, no phase difference is noticed in between the field and current.

## 4 Sheath *LR-CR* circuital description

It may be interesting to comprehend that the plasma sheath can conditionally behave as a *CR* circuit or an *LR* circuit under some special circumstances instead of as a coupled series *LCR* circuit. For a considerably larger ion transit time ($\tau_{io}$) through a wider sheath, the sheath capacitance value ($C_{sh} \propto \tau_{io}^{-1}$) significantly reduces in a considered plasma configuration. In contrast, the sheath inductance ($L_{sh} \propto \tau_{io}^3$) and sheath resistance ($R_{sh} \propto \tau_{io}^2$) values rise. It makes the sheath effectively behave as an *LR* circuit. However, for a smaller $\tau_{io}$-value, the $L_{sh}$-magnitude gets minimized, resulting in a conditional magnitude enhancement of $C_{sh}$ and $R_{sh}$. It prompts the sheath to act as an effective *CR* circuit. The current variation across the sheath during the special *CR* and *LR* circuit conditions can be evaluated by using the $L_{sh}$-, $C_{sh}$-, and $R_{sh}$-expressions (Eqs. (1)-(3)) in the Kirchhoff current-voltage laws (KCL-KVL) of electronic network analysis in a modified closed form.

Using the above network analysis, the transient current expressions for a *CR*-equivalent sheath circuit ($I_{c,CR}$) and for an *LR*-equivalent sheath circuit ($I_{c,LR}$) during their formation (charging) are respectively given [30] as

$$I_{c,CR}(t) = \left(\frac{V_o}{R_{sh}}\right) \exp\left(-\frac{t}{C_{sh}R_{sh}}\right), \tag{4}$$

$$I_{c,LR}(t) = \left(\frac{V_o}{R_{sh}}\right)\left[1 - \exp\left\{-\left(\frac{R_{sh}}{L_{sh}}\right)t\right\}\right]. \tag{5}$$

Similarly, the transient current expressions of a *CR*-equivalent sheath circuit ($I_{d,CR}$) and an *LR*-equivalent sheath circuit ($I_{d,LR}$) during their dissolution (discharging) are respectively presented in a customary symbolism [30] as

$$I_{d,CR}(t) = -\left(\frac{V_o}{R_{sh}}\right) \exp\left(-\frac{t}{C_{sh}R_{sh}}\right), \tag{6}$$

$$I_{d,LR}(t) = \left(\frac{V_o}{R_{sh}}\right) \exp\left[-\left(\frac{R_{sh}}{L_{sh}}\right)t\right]. \tag{7}$$

Here, $V_o = V(t = 0)$ denotes the initial electrostatic potential difference across the plasma sheath. Then, Eqs. (4)-(5) denote the current flow through the sheath during its formation and Eqs. (6)-(7) denote the current flow through the sheath during its dissolution. The negative nature of electric current (Eq. (6)) denotes the sheath current flowing from the sheath to the bulk plasma, when the *CR*-equivalent sheath undergoes dissolution (discharges). For the *CR*-equivalent and *LR*-equivalent sheath circuits, Eqs. (4)-(7) can be modified with the substitution of $L_{sh}$-, $C_{sh}$-, and $R_{sh}$-expressions from Eqs. (1)-(3), respectively. Application of the above mathematical expressions in Eqs. (4)-(7) results respectively in

$$I_{c,CR}(t) = \left(\frac{12\epsilon_o V_o}{c_s \lambda^2 \tau_{io}^2}\right) \exp\left[-\left(\frac{12}{\lambda^2 \tau_{io}}\right)t\right], \tag{8}$$

$$I_{c,LR}(t) = \left(\frac{12\epsilon_o V_o}{c_s \lambda^2 \tau_{io}^2}\right)\left[1 - \exp\left\{\left(\frac{40}{12\tau_{io}}\right)t\right\}\right], \tag{9}$$

$$I_{d,CR}(t) = -\left(\frac{12\epsilon_o V_o}{c_s \lambda^2 \tau_{io}^2}\right) \exp\left[-\left(\frac{12}{\lambda^2 \tau_{io}}\right)t\right], \tag{10}$$

$$I_{d,LR}(t) = \left(\frac{12\epsilon_o V_o}{c_s \lambda^2 \tau_{io}^2}\right) \exp\left[\left(\frac{40}{12\tau_{io}}\right)t\right]. \tag{11}$$

It may be noted here that, due to the special negative inductive behavior ($L_{sh} < 0$) of a plasma sheath, the negative arguments in Eq. (9) and Eq. (11) acquire positive polarity. In addition, using the previously reported plasma parameters with $V_o = 10$ V [13, 17], Eqs. (8)-(11) get simplified respectively as

$$I_{c,CR}(t) = 0.16[\exp(-1.81 \times 10^5 t)], \tag{12}$$

$$I_{c,LR}(t) = 0.16[1 - \exp(4.60 \times 10^6 t)], \tag{13}$$

$$I_{d,CR}(t) = -0.16[\exp(-1.81 \times 10^5 t)], \tag{14}$$

$$I_{c,LR}(t) = 0.16[\exp(4.60 \times 10^6 t)]. \tag{15}$$

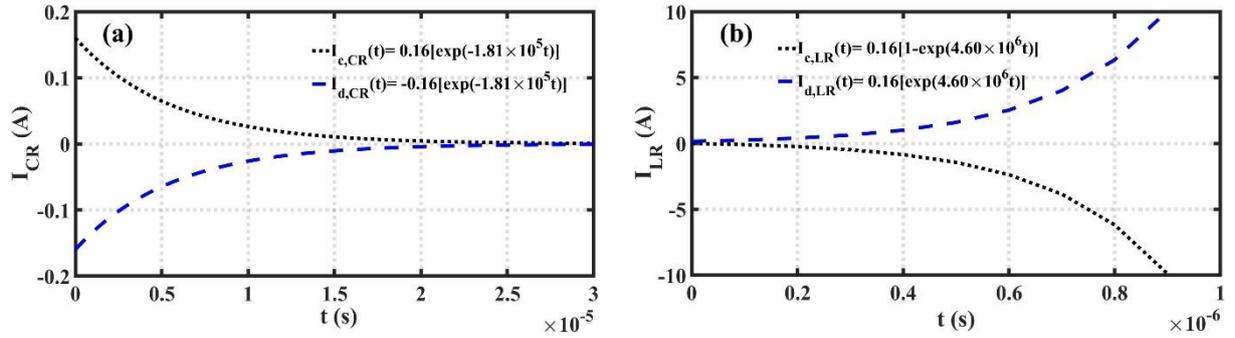

**Fig. 6** Temporal variation of sheath current in (a) *CR*-equivalent and (b) *LR*-equivalent sheath circuits during their formation (black-dotted line) and dissolution (blue-dashed line).

It may be noteworthy that the direction of the transient current flow during the sheath formation and its dissolution is opposite to each other in both *CR*- and *LR*-equivalent sheaths. It is attributable to the opposite directions of electric fields across the sheaths resulting in anti-parallel current flows. With the assumption of a plasma chamber wall or substrate located adjacent to the sheath, the positive transient currents ($I_{c,CR}, I_{d,LR} > 0$) flow towards the wall, triggering ion implantation. Therefore, we consider these wall-oriented currents as the IICs and the opposite flow of currents ($I_{d,CR}, I_{c,LR} < 0$, from sheath to bulk plasma), as usual sheath currents (USCs), since the latter diminishes in the bulk plasma, resulting in no ion implantation.

In addition, it may be interesting to notice from Fig. 6(b) that, as already mentioned before, the negative sheath inductance ($L_{sh} < 0$) makes the current continue its growth or decay without saturation. The current reversal does not occur until the plasma system is able to balance the required electric field strength across the sheath and bulk plasma. It is emphasized that the sheath current saturates during experiments, against the theoretical current predictions, once the nonlinearities take over the linear dispersive effects in diverse stages involved therein.

To study the sheath current variations with respect to the ion-transition frequency across the sheath, $\omega_{io}(= 2\pi\tau_{io}^{-1})$, we express Eqs. (8)-(11) in terms of $\omega_{io}$ in customary notations [22, 23]. The $\omega_{io}$-dependent transient current expressions across the *CR*- and *LR*-equivalent sheaths during their formation and dissolutions are respectively derived and given as

$$I_{c,CR}(t) = 1.6 \times 10^{-7} \omega_{io}^2 [\exp(-1.91 \times 10^8 \omega_{io} t)], \tag{16}$$

$$I_{c,LR}(t) = 1.6 \times 10^{-7} \omega_{io}^2 [1 - \exp(0.53 \omega_{io} t)], \tag{17}$$

$$I_{d,CR}(t) = -1.6 \times 10^{-7} \omega_{io}^2 [\exp(-1.91 \times 10^8 \omega_{io} t)], \tag{18}$$

$$I_{d,LR}(t) = 1.6 \times 10^{-7} \omega_{io}^2 [\exp(0.53 \omega_{io} t)]. \tag{19}$$

It is evident from Fig. 6 that the direction of the ion currents is opposite to each other during the formation (accumulating charges) and dissolution (dispersing charges) of the *CR*- and *LR*-equivalent sheaths. The two anti-parallel directions of current flows are as said before.

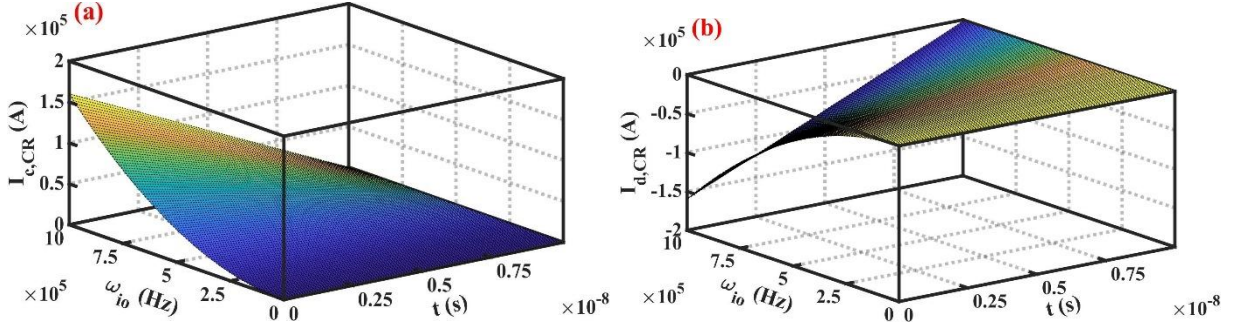

**Fig. 7** Colormap showing sheath current variation through *CR*-equivalent sheath during its (a) formation (IIC) and (b) dissolution (USC) with time ($t$) and ion-transition frequency ($\omega_{io}$).

It is evident that the transient sheath current (along z-axis) is directly proportional to the square of the $\omega_{io}$, but inversely proportional to $t$ during the *CR*-equivalent sheath formation. However, there is a saturation noticed in both $I_{c,CR}$ ($> 0$, IIC) and $I_{d,CR}$ ($< 0$, USC) with respect to $t$ and $\omega_{io}$. The opposite polarities of $I_{c,CR}$ and $I_{d,CR}$ in Fig. 7(a) and Fig. 7(b), respectively, are due to their above-mentioned anti-parallel directions. An increase in the $I_{c,CR}$-value with $\omega_{io}$ is apparently obvious as a higher rate of the ion transition across the sheath also increases its implantation on the substrate. The $I_{c,CR}$ saturation indicates the upper limit of ion implantation for that specific arrangement of the sheath parameters.

The reduction of $I_{c,CR}$ to zero with time indicates the inevitable saturation of IIC occurring at the end of the complete sheath formation. However, during the dissolution of the sheath, the $I_{d,CR}$ flowing towards the bulk plasma, also finally reaches saturation towards its zero magnitude. During the dissolution, the $\omega_{io}$ acts to make the $I_{d,CR}$ flow towards the bulk plasma instead of the plasma chamber wall or any introduced substrate. Hence, $I_{d,CR}$ apparently has no direct application as far as clearly understood.

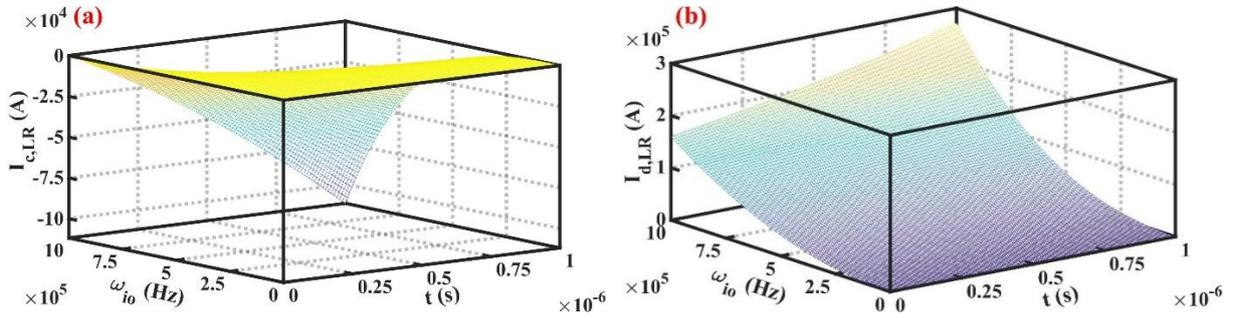

**Fig. 8** Colormap showing current variation through *LR*-equivalent sheath during its (a) formation ($I_{c,LR} < 0$, USC) and (b) dissolution ($I_{c,LR} > 0$, IIC) with respect to time ($t$) and ion-transition frequency ($\omega_{io}$).

Similarly, it may be noted that *LR*-equivalent sheath circuit current is directly proportional to both $\omega_{io}$ and $t$. The negative polarity of $I_{c,LR}$ denotes its direction of flow from the sheath to the bulk plasma during its formation. In contrast, during its dissolution, the current ($I_{d,LR} > 0$) starts to flow towards the chamber wall or introduced substrate in plasma medium. Due to the active negative induction, no current saturation is noticed. For a $L_{sh} > 0$, a saturation is likely to appear after a transient time of $t \gg L_{sh}/R_{sh}$.

It is evident from the above analyses that ion substitution, etching, thin film deposition on substrates, and other plasma-electrode interaction-based applications are possible during the *CR*-equivalent sheath formation and *LR*-equivalent sheath dissolution. In pulsed electrode plasma arrangements, such as the capacitively coupled plasmas (CCPs), both kinds of circuital sheaths are relevant as the sheath formation and dissolution occur periodically therein [32]. Therefore, regulating the sheath width with extrinsic parametric arrangements, the sheath can be made to behave either as *CR*-equivalent or *LR*-equivalent circuits in nature. An intermediate sheath width makes the sheath act as a coupled *LCR*-equivalent sheath a non-extreme case.

## 5 Results and discussions

The sheath circuital model used herein is mathematically simpler in comparison to the other established plasma sheath models, as it deals with only one governing equation. Circuital model of plasma sheath is a powerful approach that can be employed in simplifying and analyzing the behavior of the plasma sheath, especially for studying ion dynamics, voltage-current relations, and power depositions [22, 23].

The KVL-KCL equations used for circuital analysis using a linear perturbation scheme formalism justifiably ignores several variable plasma parameters. These terms are inadvertently absorbed in the circuit parametric expressions, $L_{sh}$, $C_{sh}$, and $R_{sh}$. Such absorbed parameters are the charge number density, Mach number, electric potential, etc. The assumption of constant $L_{sh}$, $C_{sh}$, and $R_{sh}$ magnitudes in the analyses [22, 23] further simplifies the numerical analysis. The spatiotemporal variation of only current perturbation evaluated from the KVL equation is sufficient to anticipate the formation of the PSW, IAW, IAW, and other relevant instabilities. This proves the higher efficacy of circuital models over other plasma sheath models, where every dependent plasma parameter must be evaluated individually to analyze the system [13].

The application of the circuital sheath model also prompts one to ponder over some special inductive, capacitive, and resistive sheath properties, typically not observed in commercial $L_{sh}$, $C_{sh}$, and $R_{sh}$ components. The plasma medium can store some significant amount of free energy in the form of electric field, magnetic field, thermal energy, etc., adequate for yielding negative $L_{sh}$, $C_{sh}$, and $R_{sh}$ values in the assumed plasma sheath circuit as illustrated before. These negative $L_{sh}$, $C_{sh}$, and $R_{sh}$ values are not solely atypical to the plasma sheath system but are also observed in diverse other non-plasmic systems as discussed below.

The negative inductance, for example, is observed in electronic filters and resonators (such as in the radio-frequency (RF) filters to shape frequency response), controlling unwanted inductance, stabilizing power supplies, etc. [33]. The negative capacitance is observed in negative capacitive transistors (to reduce power consumption), energy-efficient memory devices (to enhance charge storage, reduce leakage, improve retention time, etc.), RF and high-frequency circuits (to compensate parasitic capacitance), etc. [34]. The negative resistance is observed in oscillators (for microwaves and RF signal generation), amplifier (amplification in signal generation), high-frequency devices (e.g., radar, satellite, and communication systems), etc. [35]. In addition to the above, the circuital plasma model is also found to be relevant in the following fields with great fundamental and applied values as discussed below

(i) Ion energy modulation (IEM): The circuital plasma sheath model deals with the time-varying sheath and possible acoustic wave formations. This model can be applied in the case of commonly used CCP systems for IEM. The CCP systems use RF-driven voltage source for biasing the installed electrodes. It yields temporally fluctuating sheaths around the electrodes. These time-varying sheaths play a crucial role in shaping the ion energy distribution or IIC incident on the target substrates [36].

In the special CCPs with dual electrode voltage oscillating frequency, the interplay between the low (heavier)- and high-frequency (lighter) components can be harnessed to modulate ion energies, enabling better control over the resulting ion energy profiles [37]. Experiments incorporating a nonzero phase shift between the two driving frequencies (at two installed electrodes) further reveal that the phase angle allows for independent regulation of both ion energy and flux [38, 39].

In addition to the above, tailoring the voltage waveforms applied to the electrodes enables the generation of specific ion energy distribution (IED) profiles, which are advantageous for targeted material processing applications [40, 41]. It has been observed that increasing the fluctuation frequency of the electrode potential to very high values lead to a reduction in the peak ion energy [42–47], aiding in the identification of optimal biasing frequencies [15]. The energy modulated ions can be used for etching [48], thin film deposition [49], surface modification [50], plasma assisted lithography [51], plasma medicine [52], and so forth. The widespread applications of temporally varying sheath-driven plasma systems prove the relevance of circuital sheath model in determining the required conditions for its respective applications, expressed in terms of IIC frequency in these plasma discharge systems.

(ii) Electron heating: The time-varying sheath circuital model also holds good in electron heating in plasma discharge systems (PDSs). Various mechanisms which contribute to electron heating in such systems based on specific applications are given as follows
  (a) Ohmic (collisional) heating: This involves energy transfer from the RF electric fields produced by electrodes to the electrons via collisions with neutral atoms [53].
  (b) Stochastic (collisionless) heating: This involves energy gain of electrons from interactions with time-varying sheath induced electric fields. The time variation of the sheath is expressed in the form of rapid sheath expansion and contraction in the plasma chamber [54]. It is possible for an oscillating sheath to excite PSWs upon having an oscillation frequency $\geq$ ion plasma frequency. The mobile electrons across the sheath can be heated by the stochastic heating processes, or wave-particle interactions (Landau damping), triggered by the PSWs. It can change the overall energy distribution in the plasma system [16].
  (c) Pressure heating: It occurs from pressure gradients created by the dynamic behavior of the sheath, which transfer energy to electrons within the system [55].
  (d) Transit-time (acoustic) heating: It occurs when electrons move through spatially varying electric fields near the sheath edge, gaining energy from the wave [56].
  (e) Nonlinear and resonant heating: It involves complex electron-field interactions, including the excitation of higher harmonics and resonance phenomena that enhance energy absorption [57].

The key factors which influence electron heating include the electrode driving frequency, sheath behavior, plasma density, and operating pressure [58-64]. A circuital model can be used to determine the optimal conditions, expressed through the IIC frequency, to achieve efficient electron heating in PDSs. This proves the relevance of circuital sheath models in applied fields.

(iii) Band pass filtration: The plasma sheath at reduced inductance ($L \propto \tau_{io}^3$) condition at shorter ion transit time ($\tau_{io}$) behaves qualitatively as a *CR* circuit. The *CR* circuit equivalent sheath under this special condition can act as a band-pass filter to allow the passage of depositing ions of a specific range of frequency only, while preventing the rest. This helps in maintaining the ion deposition frequency at required magnitudes. The required deposition frequencies vary based on the introduced substrate materials. An inadequate deposition frequency can damage the substrate surface [65].

(iv) Higher harmonic generation (HHG): In CCP systems, applying a sinusoidal RF voltage causes the plasma sheaths near the electrodes to oscillate. The nonlinear behavior of these sheaths, evident in their periodic expansion and contraction, can distort the original RF waveform, producing higher harmonics (i.e., frequencies that are integer multiples of the applied RF). These harmonics significantly affect various plasma characteristics such as electron heating, ion energy distribution, and overall plasma stability, etc. [66-78]. Some key findings related to HHG from the literature are summarized below
   (a) Electric field filamentation and HHG: At very high RF driving frequencies, the electric field distribution becomes uneven, resulting in filamentation. This spatial non-uniformity boosts harmonic generation, which subsequently influences electron heating and compromises plasma uniformity [66].
   (b) Voltage- vs. current-driven discharges: Comparative studies indicate that voltage-driven CCPs produce more intense higher harmonics due to the excitation of plasma series resonance (PSR), which enhances nonlinear electron power absorption. In contrast, current-driven systems tend to suppress the PSR, leading to lower harmonic levels, proving the relevance of RF driving schemes in shaping plasma behavior [67].
   (c) Electromagnetic effects: Experiments involving electromagnetic wave propagation in CCPs show that at higher frequencies and with larger electrode areas, electromagnetic effects become more pronounced. These conditions favor higher harmonic excitation, which can interfere with plasma uniformity and reduce power deposition efficiency [68].
   (d) Influence of driving frequency: Increasing the RF bias frequency enhances the generation of higher harmonics, leading to more effective electron heating and changes in the electron energy distribution function (EEDF). This underscores the impact of RF on HHG and the control of plasma properties [69].
   (e) Nonlinear effects from biasing: Nonlinearities introduced by substrate biasing can also produce higher harmonics, influence ion energy profiles and affect material processing outcomes [67, 70].
   The higher harmonics play a vital role in practical applications such as precise control in semiconductor manufacturing, ensuring uniform plasma treatment across surfaces, and guiding the design of power delivery systems to minimize unwanted harmonic effects [71–78], and so forth.

(v) Periodic switching: The *CR* equivalent plasma sheath circuit can act as a periodic switch. The circuit may be spontaneously operated as alternating open and closed loop with temporal gaps of $\sim CR$. This adds a novel application of plasma sheath as an automated switch relevant in plasma welding and packing arrangements [79].

(vi) Harmonic electromagnetic wave generation: The sheath circuit can be used to generate relaxation oscillations of electric potential resulting in the simultaneous formations of magnetic fields, collectively forming and emanating electromagnetic waves as harmonics. The frequency (and wavelength) of the electromagnetic waves can be determined by setting the $L_{sh}$-, $C_{sh}$-, and $R_{sh}$-magnitudes, aiding the system to behave as an electromagnetic wave generator of desired frequency for practical applications [66].

(vii) Instability excitation: As mentioned above, the oscillating sheath in contact with the bulk plasma can excite diverse plasma waves, such as the IAW, electron acoustic waves, etc. The condition for these plasma wave excitations is that the oscillation amplitude must supersede (supercritical) the excitation threshold value ($\sim 10^{-2}$ J m$^{-3}$). It may happen that the excited waves take form of an instability with temporal growth of their amplitudes [17]. For instance, a nonplanar sheath and a pervading magnetic field may yield inhomogeneous fluid flow. This may trigger the shear-driven Kelvin Helmholtz instability, as observed in the astroplasmic Heliospheric regions. The velocity shear arises due to different velocities of the individual plasma layers corresponding to different plasma components [13]. This encourages extrapolating the sheath circuital model to the astroplasmic domain with required modifications, subject to investigative interests.

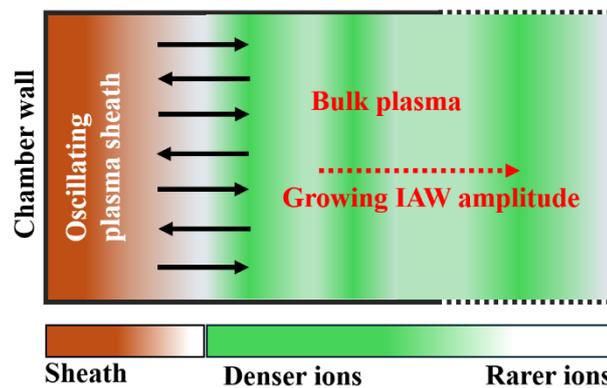

**Fig. 9** Schematic diagram showing spatiotemporal IAW amplitude growth due to the adjacent oscillating plasma sheath.

(viii) Plasma-surface interaction: Sheath oscillations in plasma chambers may increase the ion bombardment frequency and magnitude on any introduced surface in the plasma medium. This expedites applications like the ion-assisted etching or thin-film deposition through controlled plasma-sheath interactions. Uncontrolled sheath oscillations, however, may cause erosion and damage to the material. It is noticed, for example, that for sheaths of potential $\sim 100$ V in high-density plasmas, localized RF power deposition can reach to a depth of material damage [18]. Therefore, learning plasma-surface interaction and PSWs through various models has profound applications in applied physics [48, 49]. It is noteworthy that the circuital model deals with the concerned IIC responsible for the plasma-surface interaction and deposition, proving the relevance of the circuital model.

(ix) Signal distortion and analysis: The plasma sheath oscillations developed around charged antennas installed in electrodynamic tethers and space stations in astroplasmic and space environments may distort their transmitted signal due to background noise and intermittency. The electromagnetic (EM) signal harmonics originated due to some nonlinearities may make the signal fuzzy and perplexed for signal decoding analysis [21]. The circuital model may be helpful to simplify the complex sheath characteristic analysis with certain modifications in sophisticated practical circumstances.

(x) Non-invasive plasma diagnosis: Analysis of electromagnetic properties of the plasma sheath using circuital models can be helpful in extracting information regarding various plasma parameters without using any kind of conventional plasma probes directly. This analysis promotes our understanding of wave attenuation and density variation preserving the integrity of the plasma system with better efficiency, which, otherwise, gets compromised with usual sheath-enveloped probes [20]. Moreover, in the case of electrostatic wave-sheath interactions, the number density variation and sheath oscillations are found to be temporally varying and inextricably related for collective plasma analysis [76].

(xi) Plasma Sheath wave (PSW) excitation: One of the most distinctive outcomes of the circuital model is the analysis of PSWs. This model can inherently figure out the required conditions of the PSW excitation in terms of the temporal variation of current perturbation. This model analysis demonstrates that, due to the sheath width fluctuation, there is a resultant current fluctuation in the system, and vice versa. The sheath width fluctuation triggers the PSWs due to the spatiotemporal variation of the electric field across it. The excited PSWs manifested in the form of density inhomogeneities and fluctuations in the adjacent sheath region provides energy to the bulk plasma region. The energy gets transported across the bulk plasma through the IAW excitation and propagation as already proven experimentally [77].

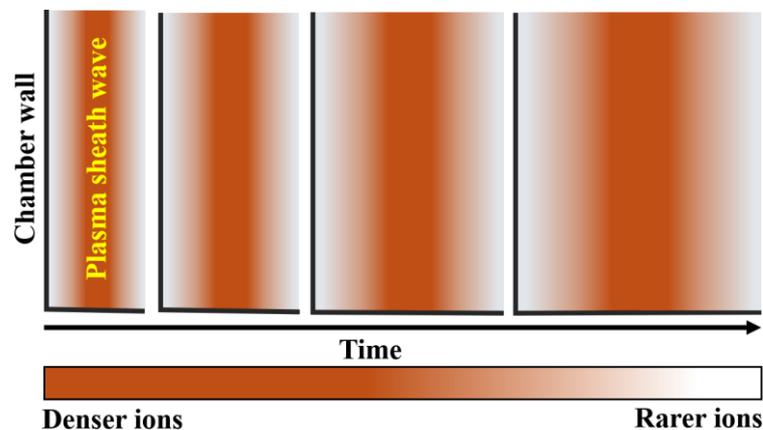

**Fig. 10** Schematic diagram showing different stages of the temporal evolution of the PSW amplitude resulting in the growing PSW instability.

(xii) Astrophysical circuital model: As highlighted before, the circuital model is also relevant across the plasmic environments which enable the excitation of DLs, solitons, vortices, and other nonlinear structures. These structures are found to excite in both laboratory and astroplasmic scenarios. Especially, in the astrophysical scenarios, the DLs are responsible for intra-circuital current flow due to the potential difference generated across the two layers of the DL. The conventional magnetofluid models of plasma are not compatible with accounting for the formation of DL structures and corresponding intra-circuital current flow. As a result, the circuital model is used to analyze the excitation of such astrophysical structures irrespective of the environmental parametric differences [78].

It is also interesting to note that the formation of DL and applied circuital model is also relevant in analyzing the energization of auroral particles (since auroras behave as electrical discharge with charge gradient), stellar flares, magnetic substorms, internal ionization in comets, and intergalactic double radio sources, etc. Furthermore, the atmospheric filaments in solar surfaces, in Venus's ionosphere, in cometary tails, in interstellar nebulae, and so forth can also be studied as DL-driven phenomena. It is noteworthy that explosive bursts of x-rays and $\gamma$-rays noticed in astrophysical and space environments result due to the explosion of the naturally developed DLs. It hereby proves the relevance of astroplasmic circuital models in diverse realistic astrophysical and space environments in the affine context of particle energization and acceleration [78].

Apart from the DLs, the solitons formed in the astroplasmas are also possible to be analyzed using the circuital model. The localized solitonic potential structures which resemble laboratory plasma sheaths enable the application of modified circuital model. The spatially rigid potential and field variations across a soliton is found suitable for the circuital analysis.

To justify the above, two plots showing the spatial variation of electric potential for DLs and solitons are presented below. The electric potential expression for the DL is $\phi_{DL}(x) = -\phi_s(1 - x/d)^{4/3}$; here $\phi_s$, $d$, and $x$ denote the total DL potential drop, DL width, and the observational variable distance, respectively [13]. The electric potential expression for the soliton is $\phi_{sol}(x) = \phi_o \operatorname{sech}^2\{(x - v_s t)/d\}$; here $v_s$, $t$, $d$, and $x$ denote soliton velocity, time of observation, soliton width, and location of observation, respectively [13]. The non-zero electric fields across these localized potential structures yield intra-structural current flow pointing at a possible application of circuital model.

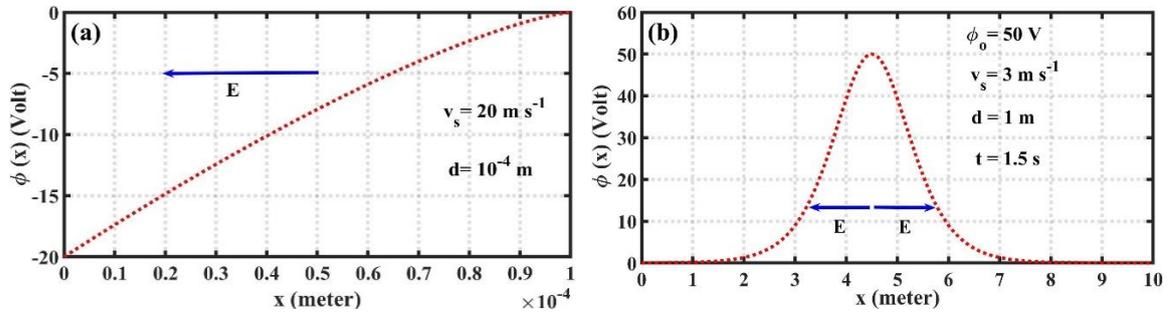

**Fig. 11** Spatial variation of typical electric potential across (a) DL and (b) soliton.

(xiii) Sheath nonlinearity: In the small order sheath oscillations, the disturbance generated is linear in nature, therefore the sinusoidally varying plasma waves occur therein. However, the sheath edge oscillation amplitude may get larger leading to nonlinear effects, such as harmonic generation, frequency mixing, etc. In the nonlinear scenarios, parametric perturbations start to influence each other, making the sheath dynamics very complicated [19]. A nonlinear analysis of the circuital sheath model can reduce the analytic complexity as the governing KVL equation deals with only the current perturbation term and rest of the parameters are absorbed within the explicit $L_{sh}$, $C_{sh}$, and $R_{sh}$ expressions.

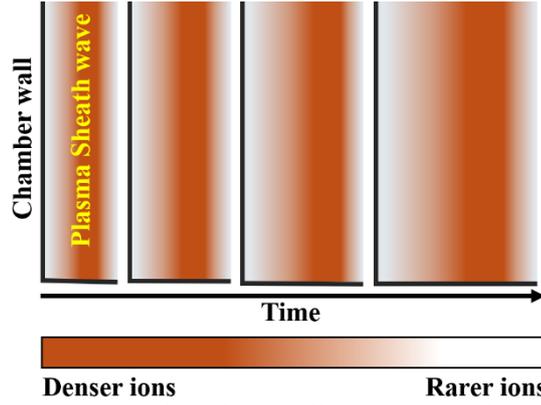
**Fig. 12** Schematic diagram portraying the nonlinear steepening of the PSW amplitude.

## 6 Summarizing and concluding remarks

In summary, it is concluded that the development of circuital sheath models is undoubtedly a remarkable achievement against other already established plasma sheath models reported so far in the literature [80-85]. In fact, the mathematical simplicity of the circuital sheath model proves it to be more convenient and efficacious for sheath-wall interaction analyses. The applications of the novel circuital plasma model are ubiquitous not only in laboratory and astrophysical plasmas, but also in electronics, communication engineering, technology, etc. [86-88]. This is the fundamental motivation for the circuital plasma sheath model analysis, as it not only prompts to study unconventional possible outcomes (such as $L_{sh}, C_{sh}, R_{sh} < 0$) but also connects plasma physics with other branches of science, adding a great value to the scientific horizon of plasma science and technology. In addition, the advantages of circuital modelling of the RF sheaths may be found in diverse directions: (i) simulation of ion energy distribution at the wall, (ii) estimation of sheath thickness and potential, (iii) calculations of ion impact energy on substrate surfaces (important for etching, sputtering), (iv) interface with external RF power sources [22, 23, 26, 28], etc.

As a future scope, the circuital model can be refined further by the inclusion of diodes [89], op-amps [90], transistors [91] in the analyses to modify the applied voltage waveform across the installed electrodes. The current decay in that case would be more complex and various other relevant plasma applications can be modelled. Furthermore, a BJT [92] or MOSFET-based [93] circuital model can account for the avalanche breakdown or negative resistance effects in certain sheath regimes for ion-rich matrix sheaths. These modified models can be used for ion implantation by the biasing of the substrate with a high amplitude pulse, such that an avalanche of ions are created and impinged on the substrate. In addition, in RF plasma systems, the sheath behaves like nonlinear time-varying capacitors [94], and they can be analyzed considering the modified diode-capacitor model to investigate the (i) voltage distribution between bulk plasma and sheath [95], (ii) sheath impedance [96], (iii) various inherent nonlinearities [97], and so forth.

The plasma sheath circuital model could be extended appropriately to model and describe astrophysical solar plasmas with the systematic inclusion of the non-local self-gravity action. The equilibrium solar plasmas are characterized by a special gravitationally and electrostatically coupled sheath, termed as the gravito-electrostatic sheath (GES) [98]. The GES structure forms at a radial balance point due to the coupling between the electrostatic (outward pushing) and self-gravitational (inward pulling) forces acting on the solar plasma system [99]. The solar surface boundary (SSB), formed due to the complex gravito-electrostatic interplay, bifurcates the entire solar plasma into two constitutive concentric parts. These are the internally self-gravitating solar interior and externally gravitating solar wind plasmas.

The GES interaction occurs because of the non-overlapping gravito-thermal coupling between the constitutive lighter electrons and heavier ions ($\Gamma_e/\Gamma_i = m_i T_e / m_e T_i \sim 10^4$, for $\Gamma_i \approx 1$, $m_i/m_e = 8.74 \times 10^3$, $T_e = 10^6$ K, and $T_i = 10^5$ K [98]). It enables the constitutive electrons to escape via the SSB more easily than the ions. Consequently, an electric charge imbalance and an effective electrostatic force of repulsion results. This electric force eventually gets counter-balanced by the attractive self-gravitational force. The interplay between these two long-range forces develops the GES and its interaction with the solar interior plasmic constituents results in outflowing supersonic solar wind streams [100, 101].

It is admissible herein that the GES formalism has several facts and faults of intrinsic origin. This model succeeds in exploring how supersonic solar wind plasmas are emitted from the SSB with its subsonic origin in the solar interior [100, 101]. A proper development of a circuital solar model could be reliably used for analyzing diverse equilibrium properties of the obscure solar plasmic structures, intercoupled through the GES action, in a futuristic astro-electronic direction of nontrivial application [102].

**Appendix-A: Comparison between circuital and non-circuital components**

The plasma circuital components are interestingly seen to have several similarities in terms of plasma multi-parametric characteristics with commercial circuital components. However, these electrical circuital components also have some contrasting differences against the commercial ones. The two distinct varieties of the circuital components are contrasted in Table A1.

**Table A1: Plasma vs. non-plasma circuital components**

| S. No. | Item | Plasma circuital components | Commercial circuital components |
|---|---|---|---|
| 1 | Formation mechanism | $L_{sh}$: Intrinsic electric field fluctuation, $C_{sh}$: Charge separation, $R_{sh}$: Resistive positive cloud in negative wall-induced ionic flow [22, 23] | $L$: Winding wires, $C$: Parallel conducting plates, $R$: Elements with high resistivity [29, 31] |
| 2 | Polarity | Negative and positive [22, 23] | Positive [29, 31] |
| 3 | Influence of free energy | Yes [22, 23] | No |
| 4 | Current nature | Ion deposition current [22] | Conduction current (for $L$ and $R$) and displacement current (for $C$) [29, 31] |
| 5 | Resistance nature | Inertial [22, 23] | Ohmic |
| 6 | Variation of component values | No (with biasing voltage) [22] | Yes |
| 7 | Resonance frequency | Variable [22, 23] | Constant |
| 8 | Formation | Spontaneous [22, 23] | Non-spontaneous |
| 9 | Physical mobility | Immobile | Mobile |
| 10 | Approximate size | $10^{-4}$ m [22, 23] | $10^{-2} - 10^{-1}$ m |
| 11 | Commercial availability | Very less | Ubiquitous |
| 12 | Commercial applicability | Less | More |
| 13 | Astrophysical relevance | Yes [78] | Not known |

**Appendix-B: Sheath model distinctions**

It is already mentioned in the text above that there are several similarities and dissimilarities between the circuital plasma model and other conventional plasma models. It hereby indicates the facts and faults of the circuital plasma model. A tabulated comparison of these different models is distinctively presented below in terms of various relevant items.

**Table B1: Distinctions between circuital and non-circuital plasma models**

| S. No. | Item | Circuital model | Non-circuital models |
|---|---|---|---|
| 1 | Plasma composition | Not considered individually [22, 23] | Considered individually [13] |
| 2 | Major results | Ion deposition current, sustaining voltage, dissipated power, and space potential [24] | Potential, electric field, velocity, charge number density [13] |
| 3 | Governing equation(s) | Kirchhof law [22, 23] | Continuity equation, momentum equation, Poisson equation, Navier-Stoke equation, etc. [13] |
| 4 | Analytical accuracy | More in relatively low-temperature plasmas [24] | More in intermediate temperature plasmas [13] |
| 5 | Computational cost/time | Less | More |
| 6 | Child law | Does not hold [22, 23] | Holds [13] |
| 7 | Sheath type | Child [22, 23] | Child and Matrix [22] |
| 8 | Sheath composition | Ions [22, 23] | Electrons, ions, dust, etc. [13] |
| 9 | Collision dominance in sheath | No [22] | Yes [22] |
| 10 | Sheath width | Variable [22] | Fixed [13] |
| 11 | Sheath electric field | Undulating [22] | Unidirectional [13] |
| 12 | Source generative field | AC [22, 23] | DC |
| 13 | Exploration | Less | More |
| 14 | Applicability | Plasma electronic industry [22] | Technological plasma processing [13] |
| 15 | Closing sheath equation | KCL and KVL [22] | Poisson equation |
| 16 | Relaxation time | Relevant (~1 μ s) [22] | Irrelevant |
| 17 | Resonance time | Relevant (~10 μ s) [22] | Irrelevant |

**Acknowledgements** The authors acknowledge the fellow researchers of Astrophysical Plasma and Nonlinear Dynamics Research Laboratory (APNDRL), Department of Physics, Tezpur University, for their active help while accomplishing this review work scientifically.

**Author contributions** All the authors have contributed equally in preparing this manuscript.

**Funding** This work has received no funding.

**Data availability** All data generated or analyzed during this study are included in this published article.